\pdfoutput=1
\RequirePackage{ifpdf}
\ifpdf % We are running pdfTeX in pdf mode
\documentclass[pdftex]{sigma}
\else
\documentclass{sigma}
\fi

\usepackage{cite}
\numberwithin{equation}{section}

\begin{document}

\allowdisplaybreaks

\renewcommand{\PaperNumber}{019}

\FirstPageHeading

\ShortArticleName{The Unruh Ef\/fect in General Boundary Quantum Field Theory}

\ArticleName{The Unruh Ef\/fect in General Boundary\\
Quantum Field Theory}

\Author{Daniele COLOSI~$^\dag$ and Dennis R\"ATZEL~$^\ddag$}

\AuthorNameForHeading{D.~Colosi and D.~R\"atzel}

\Address{$^\dag$~Centro de Ciencias Matem\'aticas, Universidad Nacional Aut\'onoma de M\'exico,
\\
\hphantom{$^\dag$}~Campus Morelia, C.P.~58190, Morelia, Michoac\'an, Mexico}
\EmailD{\href{mailto:colosi@matmor.unam.mx}{colosi@matmor.unam.mx}}

\Address{$^\ddag$~Albert Einstein Institute, Max Planck Institute for Gravitational Physics,
\\
\hphantom{$^\ddag$}~Am M\"uhlenberg 1, 14476 Golm, Germany}
\EmailD{\href{mailto:dennis.raetzel@aei.mpg.de}{dennis.raetzel@aei.mpg.de}}

\ArticleDates{Received November 07, 2012, in f\/inal form February 24, 2013; Published online March 02, 2013}

\Abstract{In the framework of the general boundary formulation (GBF) of scalar quantum f\/ield
theory we obtain a~coincidence of expectation values of local observables in the Minkowski vacuum
and in a~particular state in Rindler space.
This coincidence could be seen as a~consequence of the identif\/ication of the Minkowski vacuum as
a~thermal state in Rindler space usually associated with the Unruh ef\/fect.
However, we underline the dif\/f\/iculty in making this identif\/ication in the GBF.
Beside the Feynman quantization prescription for observables that we use to derive the coincidence
of expectation values, we investigate an alternative quantization prescription called
Berezin--Toeplitz quantization prescription, and we f\/ind that the coincidence of expectation
values does not exist for the latter.}


\Classification{81T20}

\section{Introduction}

In his seminal paper~\cite{Oe:GBQFT} Oeckl proposed an axiomatic framework for the quantum theory
that allows to formulate quantum f\/ield theory on general spacetime regions with general
boundaries.
This new formulation, named the general boundary formulation (GBF), assumes as an important
ingredient (a particular version of) the mathematical framework of topological quantum f\/ield
theory~\cite{Oe:GBQFT,Oe:KGtl,Oe:hol,Oe:aff}.
In particular the set of axioms that def\/ines the GBF implements an assignment of algebraic
structures to geometrical structures and guaranties the consistency of such an assignment.
The physical interpretation relies on a~generalization of the Born's rule to extract probabilities
from the algebraic structures (in particular amplitudes and observable amplitudes discussed below).

The main motivation at the basis of the development of the GBF is the desire to render the
formulation of quantum theory compatible with the symmetries of general relativity, in view of
a~possible future formulation of a~quantum theory of gravity.
From such perspective the GBF appears to be particularly advantageous with respect to the standard
formulation of quantum theory since it does not require a~spacetime metric for its formulation.
Indeed the axioms of the GBF necessitate only a~topological structure, and not a~metric one.
Moreover, it is worth noting that no (space)time notion enters in the def\/inition of the
generalized Born's rule.
See~\cite{Oeckl:2003vu} and \cite{Co:GBF-QG} for more details on the relevance of the GBF for the
problem of quantum gravity.

\looseness=-1
Although, as said, the spacetime background metric does not play any fundamental role in the GBF,
evidently a~general boundary quantum theory can be implemented for studying the dynamics of
f\/ields def\/ined on a~spacetime with a~def\/inite metric background.
In that case the versatility of the GBF makes it possible to consider not only initial and f\/inal
data on Cauchy surfaces as in the standard approach to quantum f\/ield theory but also the dynamics
that take place in more general regions; the main interest will be represented by compact spacetime
regions, namely regions whose boundaries have spacelike as well as timelike parts.
A certain number of
results,~\cite{Oeckl2005,Oe:2dqy,CoOe:letter,CoOe:Smatrix,CoOe:2d,Co:vac,Co:letter,Co:dS}, have
been obtained by applying the GBF for f\/ields in Minkowski space and curved spaces, among which we
cite a~new perspective on properties of the standard $S$-matrix (in particular the crossing
symmetry, that becomes a~derived property within the GBF) and the proposals of new quantization
schemes that allow a~generalization of the standard $S$-matrix.
In particular in anti-de~Sitter space, where the lack of temporal asymptotic regions obstructs the
application of the traditional $S$-matrix techniques, involving temporal asymptotic \textit{in}
and \textit{out} states, \textit{spatial asymptotic states} have been rigorously def\/ined within
the GBF of a~scalar quantum f\/ield theory and the corresponding amplitude that has been computed
for these states can then be interpreted as a~generalization of the standard
$S$-matrix~\cite{CoDoOe:AdS}.

In this article, we will investigate a~certain relation between the QFT in Minkowski space and the
QFT in Rindler space in the framework of the GBF.
The investigation is inspired by the Unruh ef\/fect which is understood as a~particular relation
between the notions of particle state in Minkowski and in Rindler spacetime.
Rindler space is the spacetime naturally associated with uniformly accelerated observers and is
isomorphic to a~submanifold of Minkowski spacetime
called the Rindler wedge.
Then, the Unruh ef\/fect can be stated as follows: linearly uniformly accelerated observers
perceive the Minkowski vacuum state (i.e.\
the no-particle state of inertial observers) as a~mixed particle state described by a~density
matrix at temperature\footnote{We set $c=\hbar=1$.} $T=a/(2\pi k_{\rm B})$, $a$ being the constant
acceleration of the observer and $k_{\rm B}$ the Boltzmann constant.
This ef\/fect was proposed by Unruh in 1976~\cite{Un:effect} and has received a~considerably
amount of attention in the literature because of its relation to other ef\/fects, like the particle
creation from black holes (the so called Hawking ef\/fect) and cosmological horizons.
Although, there are some proposals around aiming at an experimental detection of the Unruh
ef\/fect~\cite{Cri:Unruh} it was not discovered, yet.
There exist only experimental results that can be interpreted as hints to the Unruh
ef\/fect~\cite{Bell1983,Matsas1999}.

The Unruh ef\/fect must be distinguished from the result that a~uniformly accelerated Unruh--DeWitt
detector responds as if submersed in a~thermal bath when interacting with a~quantum f\/ield in the
Minkowski vacuum state~\cite{Un:effect}.
Instead, the foundation of the Unruh ef\/fect is the statement that the Minkowski vacuum state can
be interpreted as a~thermal state in Rindler space when restricted to the right Rindler wedge.
The derivation of this identity in the standard formulation of QFT is done by f\/irst, identifying
formally the vacuum state in Minkowski with an entangled state containing linear combinations of
products of $n$-particle states of the f\/ield def\/ined in the left and right Rindler wedges where
the left Rindler wedge is the point ref\/lection of the right Rindler wedge at the origin of
Minkowski space.
Then tracing out the degrees of freedom in the left Rindler wedge leads to a~density operator in
the right Rindler wedge describing a~mixed thermal state at the Unruh
temperature~\cite{Un:effect,UnWa:AccObs,ScCaDe:1981,Ta:1986,GiFr:1987,Cri:Unruh,Brout:1995rd}.

The derivation of the Unruh ef\/fect in algebraic QFT is much more sophisticated but works
primarily along the same line of argument~\cite{Sewell1982}: It is proven that the restriction of
the Minkowski vacuum state to the right Rindler wedge is identical to a~certain abstract thermal
state $\psi$ in Rindler space def\/ined as a~state fulf\/illing the KMS-condition formally given as
the identity of expectation values $\langle A(\tau)B\rangle_\psi=\langle B
A(\tau+i\beta)\rangle_\psi$, for all observables $A$, $B$ where $A(\tau)$ is the time translation
of $A$~\cite{Sewell1982}.
Sometimes, the particle content of the KMS-state is investigated by expressing it as a~density
matrix in Rindler space which is well known to be only approximately possible (see \cite[Section~5.1]{Wa:QFTCS} and  \cite[Section~6]{Earman2011})\footnote{For the problems arising in the
derivation of the Unruh ef\/fect in algebraic QFT we refer the interested reader to the article by
Earman~\cite{Earman2011}.}.

The mathematical foundation of the derivation in the standard formulation of QFT was criticized by
Narozhnyi et~al.\
in~\cite{Bel1997,Fedotov:1999gp,Narozhnyi:2000rh,Bel2001,Bel:QF-BH} which led to an answer by
Fulling and Unruh in~\cite{Fulling2004} and a~reply by Narozhnyi et~al.\
in~\cite{Narozhny2004}.
The central point of the criticism by Narozhnyi et~al.\
is that in the derivation of the Unruh ef\/fect in the standard formulation of QFT
a certain term in the mode expansion of the scalar f\/ield is neglected which is equivalent to the
requirement of an additional boundary condition at the origin of Minkowski space leading to
a~topological dif\/ferent spacetime.

We will f\/ind that also in the GBF a~relation between the Minkowski vacuum and a~state in Rindler
space cannot be derived in a~mathematical rigorous way using global mode expansions.
Moreover, since a~clear and well def\/ined notion of KMS states has not yet been implemented within
the GBF, we will not explore the relation between the Minkowski vacuum state and a~thermal state in
Rindler.
Our strategy will be to compute the expextation value of local observables for the QFTs in
Minkowski and Rindler.
In particular we will derive the coincidence of expectation values of local observables (i.e.\
observables with compact spacetime support) obtained in two dif\/ferent settings: in the f\/irst
setting expectation values are computed on the vacuum state in Minkowski spacetime and in the
second setting they are computed on a~state in Rindler spacetime that corresponds to the thermal
state known from the derivation of the Unruh ef\/fect in the standard formulation of quantum
f\/ield theory~\cite{Cri:Unruh}.

The approach used in this paper is similar in some technical aspects to the one of Unruh
and Weiss~\cite{Unruh1984}, and also presents some similarity with the derivation of the Unruh
ef\/fect in algebraic quantum f\/ield theory which is based on the restriction of the algebra of
observables to the interior of the Rindler wedge~\cite{Sewell1982}.
We emphasize that from a~physical perspective the restriction of the set of observables seen by the
Rindler observer to the interior of the Rindler wedge is a~reasonable condition since no observer
should be able to measure at its spacelike inf\/inity.

The coincidence of expectation values we will derive in this article is a~purely mathematical
result in the framework of the GBF.
Such a~result suggests a~relation between the quantum theory in Minkowski space and the one in
Rindler space which could be interpreted as a~manifestation of the Unruh ef\/fect within the GBF.
For the derivation the coincidence of expectation values, we will use a~particular quantization
prescription for observables known as Feynman quantization prescription.
It was introduced in the GBF framework in~\cite{Oe:SFobs}.
For another quantization prescription called Berezin--Toeplitz quantization prescription, we will
f\/ind no such coincidence which can be seen as a~shortcoming of this particular quantization
prescription.

Another point that must be considered in the context of the Unruh ef\/fect is the issue of the
def\/inition of temperature.
An up-to-date review on this issue can found in~\cite{Buchholz2012} where also conclusions for the
interpretation of the Unruh ef\/fect are drawn.
We shall elaborate on the notion of temperature in the context of the GBF in a~future article.

The paper is structured as follows: In Section~\ref{sec:GBF} we present a~compact review of the GBF
by specifying the two dif\/ferent representations so far implemented within the GBF, namely the
Schr\"odinger representation in which the quantum states of the f\/ield are wave functionals of
f\/ield conf\/igurations and the holomorphic representation where the states are holomorphic
functions on germs of solutions to the f\/ield equations.
In Sections~\ref{sec:Min} and \ref{sec:Rin} we formulate the general boundary quantum f\/ield
theory on Minkowski and Rindler spacetimes respectively for a~massive Klein--Gordon f\/ield both in
the Schr\"odinger and holomorphic representations.
In Section~\ref{sec:com} we show that for the massive Klein--Gordon f\/ield the GBF on Minkowski
space and the one on Rindler space are inherently dif\/ferent and cannot be compared directly using
global mode expansions.
In Section~\ref{sec:ope} we show that for observables quantized according to the Feynman
quantization prescription that are just def\/ined on the interior of the right Rindler wedge
expectation values in the Minkowski vacuum coincide with those calculated for a~state of the
quantum theory in Rindler space that corresponds to the thermal state known from the derivation of
the Unruh ef\/fect in the standard formulation of quantum f\/ield theory.
In the same section we f\/ind that this is not the case if we use the Berezin--Toeplitz
quantization prescription.
Finally, we summarize our conclusions and give an outlook in Section~\ref{sec:con}.

\section{The general boundary formulation of quantum f\/ield theory}
\label{sec:GBF}

This section presents a~short review of the two representations in which the general axioms
of~\cite{Oe:GBQFT} were implemented following the much more elaborate introduction given
in~\cite{Oe:Sch-hol}.
These are the Schr\"odinger--Feynman representation~\cite{Oe:GBQFT} and the holomorphic
representation~\cite{Oe:hol}.
We introduce the main structures that will be used in the rest of the paper, such as state spaces
and amplitude maps for both representations.

As usual, we start from an action $S[\phi]=\int_M d^Nx\,\mathcal{L}(\phi,\partial\phi,x)$ which is
considered to describe a~linear real scalar f\/ield theory in a~spacetime region $M$ of an
$N$-dimensional Lorentzian manifold $(\mathcal{M},g)$.
Denoting the
boundary\footnote{Notice that whether the boundary hypersurface $\Sigma$ is a~Cauchy surface (or
a~disjoint union of Cauchy surfaces) has no bearing on the following treatment.} of the region $M$
with $\Sigma$, we associate with this hypersurface the space $L_\Sigma$ of solutions of the
Euler--Lagrange equations (derived from the action $S[\phi]$) def\/ined in a~neighborhood of
$\Sigma$.\footnote{More precisely it is the space of germs of solutions at $\Sigma$ which is the
set of all equivalence classes of solutions where two solutions are equivalent if there exists
a~neighborhood of $\Sigma$ such that the two solutions coincide in this whole neighborhood.} The
symplectic potential on $\Sigma$ results to be
\begin{gather*}
(\theta_\Sigma)_\phi(X):=
\int_\Sigma\mathrm{d}^{N-1}\sigma\,X(x(\sigma))
\left(n^\mu\frac{\delta\mathcal{L}}{\delta\partial_\mu\phi}\right)(x(\sigma)),
\end{gather*}
where $n^{\mu}$ is the unit normal vector to $\Sigma$.
For every two elements of the space $L_\Sigma$ there is the bilinear map
$[\cdot,\cdot]_{\Sigma}:L_{\Sigma}\times L_{\Sigma}\rightarrow{\mathbb R}$ def\/ined such that
$[\xi,\eta]_{\Sigma}:=(\theta_\Sigma)_\xi(\eta)$ and the symplectic structure, that is the
anti-symmetric bilinear map $\omega_{\Sigma}:L_{\Sigma}\times L_{\Sigma}\rightarrow{\mathbb R}$
given by $\omega_{\Sigma}(\xi,\eta):=\frac{1}{2}[\xi,\eta]_{\Sigma}-\frac{1}{2}[\eta,\xi]_{\Sigma}$.
The last ingredient for the quantum theory we need to specify is a~compatible complex structure
$J_\Sigma$ represented by the linear map $J_{\Sigma}:L_{\Sigma}\rightarrow L_{\Sigma}$ such that
$J_\Sigma^2=-\text{id}$,
$\omega_{\Sigma}(J_\Sigma\cdot,J_\Sigma\cdot)=\omega_{\Sigma}(\cdot,\cdot)$
and $\omega_\Sigma(\cdot,J_\Sigma\cdot)$ is a~positive def\/inite bi-linear map.
Remark, that all ingredients but the complex structure $J_\Sigma$ are classical data uniquely
def\/ined by specifying the action.

These basic ingredients can now be used in dif\/ferent ways to specify the Hilbert spaces
associated with the boundary hypersurface $\Sigma$.
In the following subsection, we introduce the two representations developed so far within the GBF,
namely the Schr\"odinger representation, usual\-ly associated with the Feynman path integral
quantization prescription, and the holomorphic representation.

\subsection[The Schr\"odinger-Feynman representation]{The Schr\"odinger--Feynman representation}
\label{sec:Schroedinger}

In this representation, quantum states are represented by wave functionals of f\/ield
conf\/igurations.
For its implementation, it is convenient to introduce subspaces of the space $L_{\Sigma}$ of
solutions in a~neighborhood of the hypersurface $\Sigma$.
We start be def\/ining what plays the role of the ``space of momentum'', denoted by $M_\Sigma\subset L_\Sigma$,
\begin{gather*}
M_\Sigma:=\{\eta\in L_\Sigma:[\xi,\eta]=0 \ \forall\, \xi\in L_\Sigma\}.
\end{gather*}
It can be shown that $M_\Sigma$ is a~Lagrangian subspace of $L_\Sigma$.\footnote{It is this
subspace $M_\Sigma$ that def\/ines the Schr\"odinger polarization of the prequantum Hilbert space
constructed from~$L_\Sigma$, see~\cite{Oe:Sch-hol} for details.} Next, we consider the quotient
space
$Q_\Sigma:=L_\Sigma/M_\Sigma$ which corresponds the space of all f\/ield conf\/igurations on
$\Sigma$.
We denote the quotient map $L_\Sigma\rightarrow Q_\Sigma$ by $q_\Sigma$.
The last ingredient needed for the Schr\"odinger representation is the bilinear map
\begin{gather*}
\Omega_\Sigma:
\
Q_\Sigma \times Q_\Sigma \rightarrow \mathbb{C},
\qquad
(\varphi,\varphi')\mapsto 2\omega_\Sigma(j_\Sigma(\varphi),J_\Sigma
j_\Sigma(\varphi'))-\mathrm{i}[j_\Sigma(\varphi),\varphi']_\Sigma,
\end{gather*}
where $j_\Sigma$ is the unique linear map $Q_\Sigma\rightarrow L_\Sigma$ such that $q_\Sigma\circ
j_\Sigma=\text{id}_{Q_\Sigma}$ and $j_\Sigma(Q)\subseteq J_\Sigma M$.
The map $\Omega_\Sigma$ is here induced by the complex structure $J_\Sigma$ but can also be
considered independently as the single ingredient leading from the classical to the quantum theory
as it was done in~\cite{Oe:SFobs}.
However, it was shown in~\cite{Oe:Sch-hol} that there is a~one-to-one correspondence between
bilinear maps~$\Omega_\Sigma$ appropriate for the Schr\"odinger representation and complex
structures.

Notice that the symplectic potential $[\cdot,\cdot]_\Sigma$ is equivalently seen as a~map from
$L_\Sigma\times Q_\Sigma$ to the complex numbers.
The Hilbert space $\mathcal{H}_\Sigma^{\rm S}$ (the superscript~$\rm S$ refers to the Schr\"odinger
representation) is now def\/ined as the closure of the set of all coherent states
\begin{gather*}
K^{\rm S}_\xi(\varphi)=
\exp\left(\Omega_\Sigma(q_\Sigma(\xi),\varphi)+\mathrm{i}[\xi,\varphi]_\Sigma
-\frac{1}{2}\Omega_\Sigma(q_\Sigma(\xi),q_\Sigma(\xi))-\frac{\mathrm{i}}{2}[\xi,\xi]_\Sigma
-\frac{1}{2}\Omega_\Sigma(\varphi,\varphi)\right),
\end{gather*}
with respect to the inner product
\begin{gather*}
\langle K^{\rm S}_\xi,K^{\rm S}_{\xi'}\rangle:=
\int_{Q_\Sigma}\mathcal{D}\varphi \overline{K^{\rm S}_\xi(\varphi)} K^{\rm S}_{\xi'}(\varphi),
\end{gather*}
where the bar denotes complex conjugation.
The vacuum state $K^{\rm S}_0$ is then def\/ined as the coherent state with $\xi=0$.

So far we have def\/ined the kinematical aspects and we now pass to the dynamical ones.
Within the GBF the dynamics are encoded in an amplitude map
$\rho_M:\mathcal{H}_\Sigma^{\rm S}\rightarrow\mathbb{C}$ associated with the spacetime region~$M$.
In the Schr\"odinger representation for a~state $\psi^{\rm S}\in\mathcal{H}_\Sigma^{\rm S}$, the amplitude~$\rho_M$ is def\/ined in terms of the Feynman path integral prescription formally given
by~\cite{Oe:SFobs} (recall that~$\Sigma$ is the boundary of~$M$)
\begin{gather}
\rho_M\big(\psi^{\rm S}\big):=N_M\int_{L_M}\mathcal{D}\phi\, \psi^{\rm  S}(q_\Sigma(\phi))e^{\mathrm{i}  S [\phi]},
\label{eq:freampmap}
\end{gather}
where $L_M$ is the set of all f\/ield conf\/igurations in~$M$ that solve the Euler--Lagrange
equations and $N_M$ is the normalization constant def\/ined as
\begin{gather*}
N_M:=\int_{L^0_M}\mathcal{D}\phi\,e^{\mathrm{i} S [\phi]},
\end{gather*}
where $L^0_M$ is the set of all f\/ield conf\/igurations in $M$ that are zero on $\Sigma$.
With ``$\mathcal{D}\phi$'' we have denoted an hypothetical translation-invariant measure on $L_M$.
As it is well known, in general, no such measure exists in mathematical rigor.
However, using the mathematically well def\/ined holomorphic representation that we will present in
the next section Oeckl was able to give perfect mathematical sense to expressions
like~\eqref{eq:freampmap} in~\cite{Oe:Sch-hol}.
It is then possible to apply the generalized Born's rule~\cite{Oe:GBQFT,Oe:probgbf} to extract
probabilities for the amplitude map~$\rho_M$.

\subsection{The holomorphic representation}

From the complex structure $J_\Sigma$ we def\/ine the symmetric bilinear form
$g_\Sigma:L_\Sigma\times L_\Sigma\rightarrow{\mathbb R}$ as
\begin{gather*}
g_\Sigma(\xi,\eta):=2\omega_\Sigma(\xi,J_\Sigma\eta)
\qquad
\forall\, \xi,\eta\in L_\Sigma,
\end{gather*}
and assume that this form is positive def\/inite.
Next, we introduce the sesquilinear form
\begin{gather*}
\{\xi,\eta\}_\Sigma:=g_\Sigma(\xi,\eta)+2\mathrm{i}\omega_\Sigma(\xi,\eta)
\qquad
\forall\, \xi,\eta\in L_\Sigma.
\end{gather*}
The completion of $L_\Sigma$ with the inner product $\{\cdot,\cdot\}_\Sigma$ turns it into
a~complex Hilbert space.
The Hilbert space $\mathcal{H}^{\rm h}_\Sigma=H^2(L_\Sigma,\mathrm{d}\nu_{\Sigma}),$\footnote{To make
this mathematically precise one actually has to construct $\mathcal{H}^{\rm h}_\Sigma=H^2(\hat
L_{\Sigma},\mathrm{d}\nu_{\Sigma})$ where $\hat L_{\Sigma}$ is a~certain extension of $L_{\Sigma}$.
For more details about the construction of $\hat L_{\Sigma}$ and $d\nu_{\Sigma}$ we refer the
reader to~\cite{Oe:hol}.} namely the set of square integrable holomorphic functions on $L_\Sigma$,
is the closure of the set of all coherent states~\cite{Oe:hol}
\begin{gather*}
K^{\rm h}_\xi(\phi):=e^{\frac{1}{2}\{\xi,\phi\}},
\end{gather*}
where $\xi\in L_\Sigma$ and the closure is taken with respect to the inner product
\begin{gather*}
\langle K^{\rm h}_\xi,K^{\rm h}_{\xi'}\rangle:=
\int_{L_\Sigma}\mathrm{d}\nu_\Sigma(\phi)\,\overline{K^{\rm h}_\xi(\phi)}K^{\rm h}_{\xi'}(\phi),
\end{gather*}
where $\mathrm{d}\nu_\Sigma$ can be represented formally as
$\mathrm{d}\nu_\Sigma(\phi)=\mathrm{d}\mu_\Sigma(\phi)e^{\frac{1}{4}g_\Sigma(\phi,\phi)}$ with
a~certain translation invariant measure $\mathrm{d}\mu_\Sigma$.
The amplitude map for a~state $\psi^{\rm h}$ is def\/ined as
\begin{gather*}
\rho_M\big(\psi^{\rm h}\big):=\int_{L_{\tilde M}}\mathrm{d}\nu_{\tilde M}(\phi) \psi^{\rm h}(\phi),
\end{gather*}
where $L_{\tilde{M}}\subseteq L_\Sigma$ is the set of all global solutions on $M$ mapped to
$L_\Sigma$ by just considering the solutions in a~neighborhood of $\Sigma$.\footnote{More
precisely, global solutions are mapped to the corresponding germs at $\Sigma$.} The measure
$\mathrm{d}\nu_{\tilde M}$ is a~Gaussian probability measure constructed from the metric
$g_\Sigma$~\cite{Oe:hol}.\footnote{Again, we refer the reader to~\cite{Oe:hol} where the
constructions are given that make all the objects used here well def\/ined.
Additionally, in cite~\cite{Oe:Sch-hol} it was shown that the one-to-one correspondence between
maps~$\Omega_\Sigma$ and complex structures~$J_\Sigma$ leads also to mathematically well def\/ined
constructions for all the expressions in Section~\ref{sec:Schroedinger}.}

Independent of the representation the amplitude for coherent states turns out to be\footnote{See
equation (31) of~\cite{Oe:SFobs} for normalized coherent states and equation (43) in~\cite{Oe:hol}
as well as~\cite{Oe:Sch-hol}.}
\begin{gather}
\rho_M(K_\xi)=
\exp\left(\frac{1}{2}g_\Sigma\big(\xi^R,\xi^R\big)-\frac{1}{2}g_\Sigma\big(\xi^I,\xi^I\big)
-\frac{\mathrm{i}}{2}g_\Sigma\big(\xi^R,\xi^I\big)\right),
\label{eq:freeamphol}
\end{gather}
where $\xi^R,\xi^I\in L_{\tilde M}$ and $\xi=\xi^R+J_\Sigma\xi^I$.

\section{GBF in Minkowski and Rindler spacetimes}

We start with the action for the real massive Klein--Gordon f\/ield on $(1+1)$-dimensional Minkowski
spacetime $\mathcal{M}=(\mathbb{R}^2,\eta=\text{diag}(1,-1))$ which is given by
\begin{gather}
\label{eq:action}
S[\phi]=
\frac{1}{2}\int\mathrm{d}^2x\left(\eta^{\mu\nu}\partial_\mu\phi\partial_\nu\phi-m^2\phi^2\right).
\end{gather}
The resulting symplectic potential for a~spacetime region~$M$ with boundary hypersurface~$\Sigma$ is
\begin{gather*}
(\theta_\Sigma)_\xi(\phi)=
\frac{\epsilon}{2}\int_\Sigma\mathrm{d}\sigma\, \xi(x(\sigma))\left(n^\mu\partial_\mu\phi\right)(x(\sigma)),
\end{gather*}
with $n^\mu$ the normalized hypersurface normal vector f\/ield pointing inside the region $M$
and $\epsilon=\pm1$ if $\Sigma$ is everywhere spacelike/timelike respectively\footnote{We stick
here to the conventions used in~\cite{Oe:hol} and earlier publications.}.
The derivative $\zeta:=\frac{d}{d\sigma}\Sigma(\sigma)$ of the embedding function $\Sigma(\sigma)$
is normalized as $\eta^{\mu\nu}\zeta_\mu\zeta_\nu=1$.

\subsection{Minkowski spacetime}
\label{sec:Min}

We want to investigate the GBF in a~region $M\subset{\mathcal{M}}$ bounded by the disjoint union of
two spacelike hypersurfaces represented by two equal time hyperplanes (this corresponds to the
standard setting), which we denote as $\Sigma_{1,2}:\{t=t_{1,2}\}$, i.e.\
$M={\mathbb R}\times[t_1,t_2]$.
Then the boundary of the region~$M$ corresponds to the disjoint union
$\Sigma:=\Sigma_{1}\cup\overline\Sigma_{2}$ (the bar denotes the inverted orientation).
The set of solutions in the neighborhood of $\Sigma$ decomposes in a~direct sum as
$L_\Sigma=L_{\Sigma_1}\oplus L_{\overline{\Sigma}_2}$ where $L_{\Sigma_1}$
and $L_{\overline{\Sigma}_2}$ are the sets of solutions in the neighborhood of~$\Sigma_1$
and~$\overline{\Sigma}_2$ respectively each equipped with the corresponding symplectic form
$\omega_{\Sigma_1}$ respectively $\omega_{\overline{\Sigma}_2}$ and a~complex structure
$J_{\Sigma_1}$ respectively~$J_{\overline{\Sigma}_2}$.
The inversion of the orientation is implemented by the identif\/ication
$[\phi,\phi']_{\overline{\Sigma}_2}=-[\phi,\phi']_{\Sigma_2}$
and $J_{\overline{\Sigma}_2}=-J_{\Sigma_2}$.
The corresponding Hilbert space associated with $\Sigma$ is given by the tensor product
$\mathcal{H}_\Sigma=\mathcal{H}_{\Sigma_1}\otimes\mathcal{H}_{\overline{\Sigma}_2}$, where
$\mathcal{H}_{\Sigma_1}$ and $\mathcal{H}_{\overline{\Sigma}_2}$ are the Hilbert spaces associated
with the hypersurface $\Sigma_1$ and~$\overline{\Sigma}_2$ respectively and the inversion of the
orientation translates to the level of the Hilbert spaces by the map
$\iota:\mathcal{H}_{\Sigma_2}\rightarrow\mathcal{H}_{\overline{\Sigma}_2}$,
$\psi\mapsto\overline{\psi}$ as can be seen from the def\/inition of the coherent states in
Section~\ref{sec:GBF}.

In order to provide an explicit expression to the structures introduced in the previous section we
expand the scalar f\/ield in a~complete basis of solutions of the equation of motion,
\begin{gather}
\label{eq:parameterization}
\phi(x,t)=\int dp\,\left(\phi(p)\psi_p(x,t)+\text{c.c.}\right),
\end{gather}
where $\psi_p(x,t)$ are chosen to be the eigenfunctions of the boost generator, namely the boost
modes\footnote{It is assumed that an inf\/initely small imaginary part is added to $t$.
Moreover, the integral over $p$ in~\eqref{eq:parameterization} must be extended from $-\infty$ to
$+\infty$. Usually the expansion is given in the basis of plane wave solutions.
However, it turns out to be more convenient for our purposes to use the boost modes.}
\begin{gather}
\psi_p(x,t)=
\frac{1}{2^{3/2}\pi}
\int_{-\infty}^{\infty}\mathrm{d}q\,\exp\left(\mathrm{i}m(x \sinh q-t \cosh q)-\mathrm{i}p q\right)
\nonumber
\\
\phantom{\psi_p(x,t)}
=e^{-\mathrm{i}\omega t}\frac{1}{2^{3/2}\pi}
\int_{-\infty}^{\infty}\mathrm{d}q\,\exp\left(\mathrm{i}m x \sinh q-\mathrm{i}p q\right),
\label{eq:boostmodes}
\end{gather}
where we have introduced the operator $\omega=\sqrt{-\partial_x^2+m^2}$.
These modes are normalized as
\begin{gather}
\omega_{\Sigma_i}(\overline{\psi_p},\psi_{p'})=\delta(p-p'),
\qquad
\omega_{\Sigma_i}(\psi_p,\psi_{p'})=\omega_{\Sigma_i}\big(\overline{\psi_p},\overline{\psi_{p'}}\big)=0.
\label{eq:normboostmodes}
\end{gather}
The Hilbert space $\mathcal{H}_i$ of the quantum theory, associated to a~hyperplane $\Sigma_i$,
$i=1,2$, is def\/ined by the vacuum state written in the Schr\"odinger representation as
\begin{gather}
K_{0,\Sigma_i}^{\rm S} (\varphi_i)=
N\exp\left(-\frac{1}{2}\int\mathrm{d}x\,\varphi_i(x)(\omega\varphi_i)(x)\right),
\qquad
i=1,2,
\label{eq:vacMin}
\end{gather}
$N$ being a~normalization constant and $\varphi_i\in Q_{\Sigma_i}$ are the boundary f\/ield
conf\/igurations on the hypersurface $\Sigma_i$, namely $\varphi_i(x)=\phi(x,t)\big|_{t=t_i}$.
This vacuum state corresponds to the standard Minkowski vacuum state\footnote{In fact, the standard
plane wave basis and the basis of the boost modes are related by a~unitary transformation.}, whose
GBF expression has been given in~\cite{CoOe:Smatrix}, and it is uniquely def\/ined by the complex
structure~\cite{ashtekar1975}
\begin{gather*}
J_{\Sigma_i}=\frac{\partial_t}{\sqrt{-\partial_t^2}},
%\label{eq:cstructure}
\end{gather*}
which def\/ines a~unitary complex structure on $L_\Sigma$ in the sense that it is compatible with
the dynamics of the f\/ield.
The boost modes~\eqref{eq:boostmodes} are eigenfunctions of this complex structure, i.e.\
$J_{\Sigma_i}\psi_p=-\mathrm{i}\psi_p$.
The structures introduced in the previous section, namely the symplectic form
$\omega_{\Sigma_{i}}(\cdot,\cdot)$, the metric $g_{\Sigma_{i}}(\cdot,\cdot)$ and the inner product
$\{\cdot,\cdot\}_{\Sigma_{i}}$, evaluated for two solutions $\phi,\phi'\in L_{\Sigma_{i}}$,
$i=1,2$, take the form
\begin{gather}
\omega_{\Sigma_{i}}(\phi,\phi') =\frac{\mathrm{i}}{2}\int_{-\infty}^{\infty} \mathrm{d} p\,\left(
\overline{\phi(p)} \phi'(p) - \phi(p) \overline{\phi'(p)} \right),
\label{eq:symstrMin}
\\
g_{\Sigma_{i}}(\phi,\phi')  =\int_{-\infty}^{\infty} \mathrm{d} p\, \left( \overline{\phi(p)}
\phi'(p) + \phi(p) \overline{\phi'(p)} \right),
\label{eq:metMin}
\\
\left\{\phi,\phi'\right\}_{\Sigma_{i}}  =
g_{\Sigma_{i}}(\phi,\phi')+2\mathrm{i}\omega_{\Sigma_{i}}(\phi,\phi')=2\int_{-\infty}^{\infty} \mathrm{d}
p\, \phi(p) \overline{\phi'(p)}.
\label{eq:innproMin}
\end{gather}
The dense subset of the Hilbert space associate to~$\Sigma_i$, def\/ined by the coherent states, as
well as the amplitude map associated to the region~$M$ are implementable in terms of the above
quantities.

\subsection{Rindler spacetime}
\label{sec:Rin}

For the quantization of the scalar f\/ield in Rindler spacetime we consider again the action in
equation~\eqref{eq:action} but restricted to the right wedge of Minkowski space, namely
$\mathcal{R}:=\{(x,t)\in\mathcal{M}:t^2-x^2\leq0,\,x>0\}$, which is covered by the Rindler
coordinates $(\rho,\eta)$ such that $\rho\in{\mathbb R}^+$ and $\eta\in{\mathbb R}$.
The relation between the Cartesian coordinates $(x,t)$ and the Rindler ones is $t=\rho\sinh\eta$
and $x=\rho\sinh\eta$, and the metric of Rindler space results to be $ds^2=\rho^2d\eta^2-d\rho^2$.
We consider the region $R\subset{\mathcal{R}}$ bounded by the disjoint union of two
equal-Rindler-time hyperplanes $\Sigma_{1,2}^R:\{\eta=\eta_{1,2}\}$, i.e.\
$R={\mathbb R}^+\times[\eta_1,\eta_2]$.
In order to repeat the construct of the quantum theory implemented in Minkowski spacetime, we start
by expanding the f\/ield in a~complete set of solutions of the equation of motion,
\begin{gather*}
\phi^R(\rho,\eta)=\int_{0}^{\infty}\mathrm{d}p\left(\phi^R(p)\phi^R_p(\rho,\eta)+\text{c.c.}\right),
\end{gather*}
where the Fulling modes~\cite{Ful} $\phi^R_p$ read
\begin{gather}
\phi^R_p(\rho,\eta)=\frac{(\sinh(p\pi))^{1/2}}{\pi}K_{\mathrm{i}p}(m\rho)e^{-\mathrm{i}p\eta},
\qquad
p>0.
\label{eq:Rinmodes}
\end{gather}
$K_{\mathrm{i}p}$ is the modif\/ied Bessel function of the second kind, also known as Macdonald
function~\cite{GrRi}.
The modes~\eqref{eq:Rinmodes} are normalized as
\begin{gather*}
\omega_{\Sigma_i^R}\big(\overline{\phi^R_p},\phi^R_{p'}\big)=\delta(p-p'),
\qquad
\omega_{\Sigma_i^R}\big(\phi^R_p,\phi^R_{p'}\big)=
\omega_{\Sigma_i}\big(\overline{\phi^R_p},\overline{\phi^R_{p'}}\big)=0.
\end{gather*}
The Hilbert space associated with the hypersurface $\Sigma_i^R$, $i=1,2$, is characterized by the
fol\-lo\-wing vacuum state in the Schr\"odinger representation, expressed in terms of the boundary
f\/ield conf\/iguration~$\varphi_i$,
\begin{gather*}
K_{0,\Sigma_i^R}^{\rm S}(\varphi_i)=
N\exp\left(-\frac{1}{2}\int\frac{\mathrm{d}\rho}{\rho}\,\varphi_i(\rho)(\omega\varphi_i)(\rho)\right),
\qquad
i=1,2,
\end{gather*}
where $\omega$ now denotes the operator $\omega=\sqrt{(\rho\partial_{\rho})^2-m^2}$ and $N$ is
a~normalization factor.
This vacuum state is in correspondence with the following complex structure, def\/ined by the
derivative with respect to the Rindler time coordinate $\eta$,
\begin{gather}
J_{\Sigma_i^R}=\frac{\partial_{\eta}}{\sqrt{-\partial_{\eta}^2}},
\label{eq:cstructureRindler}
\end{gather}
and the Fulling modes~\eqref{eq:Rinmodes} are eigenfunctions of this complex structure:
$J_{\Sigma_i^R}\phi^R_p=-\mathrm{i}\phi^R_p$.
The algebraic structures def\/ined on the hypersurface $\Sigma_i^R$, considered for two solutions
$\phi^R,\psi^R\in L_{\Sigma_i^R}$ result to be
\begin{gather*}
\omega_{\Sigma_{i}^R}\big(\phi^R,\psi^R\big) =\frac{\mathrm{i}}{2}\int_{0}^{\infty} \mathrm{d} p\,\left(
\overline{\phi^R(p)} \psi^R(p) - \phi^R(p) \overline{\psi^R(p)} \right),
%\label{eq:symstrRin}
\\
g_{\Sigma_{i}^R}\big(\phi^R,\psi^R\big)  =\int_{0}^{\infty} \mathrm{d} p\,\left( \overline{\phi^R(p)}
\psi^R(p) + \phi^R(p) \overline{\psi^R(p)} \right),
%\label{eq:metRin}
\\
\left\{ \phi^R,\psi^R \right\}_{\Sigma_{i}^R}  =
g_{\Sigma_{i}^R}\big(\phi^R,\psi^R\big)+2\mathrm{i}\omega_{\Sigma_{i}^R}\big(\phi^R,\psi^R\big)=2\int_{0}^{\infty}
\mathrm{d} p\, \phi^R(p) \overline{\psi^R(p)}.
%\label{eq:innproRin}
\end{gather*}
It is important to notice that in order for the quantum theory to be well def\/ined the following
condition must be imposed on the f\/ield in Rindler space: $\phi^R(\rho=0,\eta)=0$.
Indeed the complex structure~\eqref{eq:cstructureRindler} is well def\/ined except in the origin of
Minkowski spacetime as can be seen by expressing~\eqref{eq:cstructureRindler} in terms of the
Cartesian coordinates $(x,t)$,
\begin{gather*}
J_{\Sigma_i^R}=\frac{x\partial_{t}+t\partial_x}{\sqrt{-(x\partial_{t}+t\partial_x)^2}}.
\end{gather*}
The relevance of such a~condition in the derivation of the Unruh ef\/fect has been emphasized,
and discussed both in the canonical and algebraic approach to quantum f\/ield theory, by Belinskii
et~al.\
in~\cite{Bel1997,Fedotov:1999gp,Narozhnyi:2000rh,Bel2001,Bel:QF-BH}.
This condition plays indeed a~fundamental role in the attempt to compare the quantum theories in
Minkowski and Rindler spacetimes, as will be discussed in the next section.

\section{Comparison of Minkowski and Rindler quantization \\using global mode expansions}
\label{sec:com}

The Unruh ef\/fect can be expressed as the statement that an uniformly accelerated observer
perceives the Minkowski vacuum state as a~mixed thermal state at a~temperature proportional to its
acceleration.
In many treatments of the Unruh ef\/fect this claim relies on a~comparison between the quantum
theory in Minkowski and the one in the right Rindler wedge which is naturally associated with an
accelerated observer.

The quantization scheme proposed by Unruh to implement such a~comparison rests on the properties of
particular linear combinations of the boost modes, known as the Unruh modes:
\begin{gather}%\label{eq:bogolubov}
R_p(x,t)=\frac{1}{\sqrt{2\sinh(p\pi)}}\left(e^{p\pi/2}\,\psi_p(x,t)-e^{-p\pi/2}\,\overline{\psi_{-p}(x,t)}\right),
\label{eq:URmodes}
\\
L_p(x,t)=\frac{1}{\sqrt{2\sinh(p\pi)}}\left(e^{p\pi/2}\,\overline{\psi_{-p}(x,t)}-e^{-p\pi/2}\,\psi_p(x,t)\right),
\label{eq:ULmodes}
\end{gather}
with $p>0$, whose normalization is determined by the one of the boost
modes~\eqref{eq:normboostmodes}.
The key property of the Unruh modes is their behavior when evaluated in the right and left wedge of
Minkowski spacetime\footnote{The left wedge of Minkowski spacetime is the ref\/lection of the right
wedge with respect to the origin, namely ${\mathcal L}:\{x\in\mathcal M:x^2\leq0,x<0\}$.}: the
modes $R_p(x,t)$ (respectively $L_p(x,t)$) vanish for $(x,t)\in{\mathcal L}$ (respectively
$(x,t)\in{\mathcal R}$) and moreover $R_p(x,t)$ coincide with the Fulling modes~\eqref{eq:Rinmodes}
in ${\mathcal R}$.
Equations~\eqref{eq:URmodes} and \eqref{eq:ULmodes} are then interpreted as Bogolubov
transformations connecting the expansion of the f\/ield in the basis of the boost modes and the one
in the Unruh modes.
The existence of such a~Bogolubov transformation allows to relate the corresponding annihilation
and creation operators def\/ined within the canonical approach for the two quantization schemes
and consequently the quantum states def\/ined between the quantum theories.
By inverting relations~\eqref{eq:URmodes} and \eqref{eq:ULmodes} and substituting in the expansion
of the f\/ield in the basis of the boost modes provides an expression of the f\/ield in terms of
the Unruh modes.
Then, the restriction of the f\/ield to the right Rindler wedge allows a~direct comparison, via the
Bogolubov transformation, to the expansion of the f\/ield in Rindler spacetime in the basis of the
Fulling modes.

In their review~\cite{Cri:Unruh}, Crispino et~al.\ provide the following def\/inition of the
Unruh ef\/fect (end of Section~II.B): ``The Unruh ef\/fect is def\/ined in this review as the fact
that the usual vacuum state for QFT in Minkowski spacetime restricted to the right Rindler wedge is
a~thermal state~[\dots]''.
The derivation of the Unruh ef\/fect in the standard formulation of QFT is to use the Bogolubov
transformations derived from~\eqref{eq:URmodes} and~\eqref{eq:ULmodes} to calculate the expectation value of the
canonical operator corresponding to the Rindler particle number operator evaluated on the Minkowski
vacuum state which turns out to be that of a~thermal bath of Rindler particles.
This derivation relies on the properties of particle states and number particle operator which have
a~global nature\footnote{In the canonical approach particle states are determined by the action of
the creation and annihilation operators whose def\/inition involves the values that the f\/ield
takes over a~(non compact) Cauchy surface.
In the GBF treatment, particle states are elements of the Hilbert space associated with the
hypersurface under consideration and the construction of such a~Hilbert space depends on the
f\/ield conf\/igurations on this hypersurface, i.e.\
it depends on the global properties of the elements of the space~$L_{\Sigma}$.}.

The above mentioned derivation of the Unruh ef\/fect in the standard formulation of QFT has been
criticized~\cite{Bel:QF-BH} in virtue of the existence of the boundary condition mentioned at the
end of the preceding section.
We will not repeat here the arguments presented in the cited papers but limit ourself to a~few
considerations.
First, we notice that the correct expansion of the f\/ield in Minkowski spacetime in terms of the
Unruh modes reads
\begin{gather}
\phi(x,t)=
\lim_{\epsilon\rightarrow0}\int_{\epsilon}^{\infty}\mathrm{d}p\left(r(p)R_p(x,t)+l(p)L_p(x,t)+\text{c.c.}\right)
\nonumber
\\
\phantom{\phi(x,t)=}{}
+\lim_{\epsilon\rightarrow0}\int_{-\epsilon}^{\epsilon}\mathrm{d}p\left(\phi(p)\psi_p(x,t)+\text{c.c.}\right),
\label{eq:expMinrl}
\end{gather}
which does not coincide with the expansion of the f\/ield in the basis of the Fulling modes for the
world points located inside the right Rindler wedge, i.e.\
for $(x,t)\in{\mathcal R}$.
The dif\/ference is due to presence of the last term in the r.h.s.\
of~\eqref{eq:expMinrl}: for $(x,t)\neq(0,0)$ the integral vanishes in the limit
$\epsilon\rightarrow0$ because the boost modes take f\/inite values\footnote{We are here assuming
that $|\phi(p)|<\infty$ in the limit $p\rightarrow0$.}, but for $(x,t)=(0,0)$ the contribution of
this term cannot be neglected since the boost modes reduce to a~delta function, as can be seen from
expression~\eqref{eq:boostmodes}, i.e.\
$\psi_p(0,0)=\delta(p)/\sqrt{2}$.
The importance of this term is evident when considering the algebraic structures needed for the
implementation of the GBF.
For simplicity and without loss of generality, we consider the hyperplane $\Sigma_0$ at $t=0$
and the structures~\eqref{eq:symstrMin},~\eqref{eq:metMin} and \eqref{eq:innproMin} def\/ined on it.
When restricting $\Sigma_0$ to the right wedge $\mathcal R$, by using~\eqref{eq:expMinrl} we obtain
(we denote such restriction with the superscript $(\mathcal R)$)
\begin{gather}
\omega^{(\mathcal R)}_{\Sigma_0}(\phi,\phi')=
\omega_{\Sigma^R_0}\big(\phi^R,{\phi^R}'\big)+\lim_{\epsilon\rightarrow0}\mathrm{i}
\int_{0}^{\epsilon}\mathrm{d}p\,\frac{\cosh(p\pi)}{\sinh(p\pi)}
\left[\phi(p) \overline{\phi(p)'}-\overline{\phi(p)} \phi'(p)\right],
\label{eq:symstrrest}
\end{gather}
where $\Sigma^R_0$ is the semi-hyperplane $\eta=0$ ($\eta$ being the Rindler time) corresponding to
the intersection $\Sigma_0\cap{\mathcal R}$.
Analogue expressions are obtained for the restriction of~\eqref{eq:metMin} and \eqref{eq:innproMin}.
From~\eqref{eq:symstrrest} we notice that the restriction to~${\mathcal R}$ of the symplectic
structure def\/ined in Minkowski spacetime coincides with the symplectic structure in Rindler
spacetime only if the second term in~\eqref{eq:symstrrest} vanishes, and for that we must impose
$\phi(p)=0$ for $p=0$.
However, requiring such a~condition implies imposing the vanishing of the f\/ield at the left edge
of the right wedge, namely at the origin of Minkowski spacetime.
While this is a~built-in boundary condition the f\/ield in Rindler has to satisfy\footnote{This is
usual decay condition at inf\/inity for the f\/ield in Rindler spacetime since the origin of
Minkowski spacetime corresponds to spatial inf\/inity from the point of view of a~uniformly
accelerated observer.}, there is no reason to require the same condition for the f\/ield in
Minkowski.
Indeed imposing such condition translates in the exclusion of the zero boost mode from the set of
modes on which the f\/ield is expanded, namely in the suppression of the second term
in~\eqref{eq:expMinrl}.
But the remaining expression will then represent the expansion of a~dif\/ferent f\/ield in
Minkowski spacetime, namely a~f\/ield that satisf\/ies a~zero boundary condition at $(x,t)=(0,0)$.
The appearance of this condition is a~consequence of the fact that the basis of the boost modes is
not anymore complete without the zero boost mode.
The consequence of this fact at the quantum level manifests in the loss of the translational
invariance of the Minkowski vacuum state, see~\cite{Bel:QF-BH} for a~detailed discussion on this
point.

The claim that the Minkowski vacuum state can be written as an entangled state composed by multi
particle states def\/ined in the left and right wedges is consequently not acceptable.
For example Unruh and Wald~\cite{UnWa:AccObs} provide the following equality
\begin{gather}
|0_M\rangle=\prod_j N_j\sum_{n_j=
0}^{\infty}e^{-\pi n_j\omega_j/a}|n_j,{\mathcal L}\rangle\otimes|n_j,{\mathcal R}\rangle,
\label{eq:vacstaent}
\end{gather}
where $|0_M\rangle$ is the vacuum state in Minkowski space, $N_j=(1-\exp(-2\pi\omega_j/a))^{1/2}$,
and $|n_j,{\mathcal L}\rangle$, $|n_j,{\mathcal R}\rangle$ represent the state with $n_j$ particles in
the mode $j$ in the left, right wedge ${\mathcal L}$ and ${\mathcal R}$ respectively.
From~\eqref{eq:vacstaent} it is then possible to obtain a~reduced density matrix by tracing over
the degrees of freedom in the left wedge,
\begin{gather}
\varrho^{(\mathcal R)}=\prod_j N_j^2\sum_{n_j=
0}^{\infty}e^{-2\pi n_j\omega_j/a}|n_j,{\mathcal R}\rangle\otimes\langle n_j,{\mathcal R}|.
\label{eq:rhoMinR}
\end{gather}
This reduced density matrix is interpreted as representing the restriction of the Minkowski vacuum
to the region ${\mathcal R}$.
As already mentioned, the left and right hand side of formulas~\eqref{eq:vacstaent}
and~\eqref{eq:rhoMinR} refer to states that belong to unitarily inequivalent quantum theories
and are consequently not mathematically well-def\/ined.

In fact, that~\eqref{eq:rhoMinR} can only be formally true was already observed much earlier than
the articles~\cite{Bel1997,Fedotov:1999gp,Narozhnyi:2000rh,Bel2001,Bel:QF-BH} started the
discussion about the boundary condition at the origin.
For example, in~\cite{Wa:QFTCS}, we f\/ind the statement that the Minkowski vacuum state cannot be
expressed as a~density matrix in the quantum f\/ield theory in Rindler space.
This mathematical problem was avoided by showing that the Minkowski vacuum is a~KMS-state in
Rindler space and using KMS-condition to def\/ine thermal states in a~way that does not rely on the
particle number operator~\cite{Sewell1982}.
Since there is no def\/inition of KMS-states in the GBF, yet, we cannot give a~mathematically
rigorous version of equation~\eqref{eq:rhoMinR} in this article.

However, inspired by some results derived within the algebraic approach to quantum f\/ield
theory\footnote{We refer in particular to Fell's theorem~\cite{Fell} and the work of
Verch~\cite{Verch:1992eg}.}, in the next section we present a~result, obtained within the GBF, that
suggests the existence of a~relation between the Minkowski vacuum state and a~state of the quantum
theory in Rindler spacetime that corresponds to the thermal state known from the derivation of the
Unruh ef\/fect in the standard formulation of quantum f\/ield theory.
To be more precise, we compute the expectation value of a~Weyl observable def\/ined on a~compact
spacetime region in the interior of the right Rindler wedge in two dif\/ferent contexts: f\/irst on
the vacuum state in Minkowski spacetime and then on a~state in Rindler space, whose form represents
the analogue of the r.h.s.\ of~\eqref{eq:rhoMinR} in the GBF language.
It turns out that these two expectation values are equal when the observables are quantized
according to the Feynman quantization prescription.

\section{The relation between operator amplitudes\\ on Minkowski and Rindler space}
\label{sec:ope}

In this section, we derive the coincidence of the expectation values of local observables computed
in the Minkowski vacuum and a~certain state in Rindler spacetimes which is the central result of
this article.
The observables we consider have been called Weyl observables in~\cite{Oe:SFobs} and are given by
an exponential of a~linear functional of the f\/ield,
\begin{gather}
F(\phi)=\exp\left(\mathrm{i}\int\mathrm{d}^2x\,\mu(x)\phi(x)\right),
\label{eq:Weylobs}
\end{gather}
in our case $\mu(x)$ is assumed to have compact support in the interior of the right wedge
${\mathcal R}$ and~\eqref{eq:Weylobs} is consequently a~well def\/ined observable in both Minkowski
and Rindler spacetime.
The interest for looking at the Weyl observables is twofold: f\/irst, consistent quantization
schemes have been established within the GBF, and second, general results concerning expectation
values of these observables have been obtained in~\cite{Oe:SFobs}.
Here we consider the Feynman and Berezin--Toeplitz quantizations of~\eqref{eq:Weylobs} and compute
for the corresponding quantum observables two dif\/ferent expectation values: one on the vacuum
state in Minkowski spacetime and the other on a~state in Rindler spacetime that corresponds to the
thermal state known from the derivation of the Unruh ef\/fect in the standard formulation of
quantum f\/ield theory.

In the GBF, as in algebraic quantum f\/ield theory, quantum observables $O_M$ are associated with
a~spacetime region~$M$.
They are def\/ined by a~linear map, called observable map or observable amplitude, from (a dense
subspace of) the Hilbert space associated with the boundary of the region to the complex numbers,
$O_M:\mathcal{H}_{\Sigma}\rightarrow{\mathbb C}$, $\Sigma$ being the boundary of the region~$M$.
A set of axioms establishes the properties of this map, in particular the spacetime composition of
observables.
The specif\/ic form of the observable map depends on the quantization scheme adopted.
In the following two sections, the Feynman and Berezin--Toeplitz quantization schemes combined with
the Schr\"odinger and holomorphic representations are used for the Weyl observable in the settings
specif\/ied above.

The Feynman quantization prescription is inspired from the purely formal expression for the
operator amplitude known from the path integral formulation of quantum f\/ield theory.
The corresponding observable map associated with an observable~$O_M$ evaluated on a~state
$\psi^{\rm S}\in\mathcal{H}_{\Sigma}^{\rm S}$ in the Schr\"odinger representation takes the form
\begin{gather*}
\rho^{O_M}_M\big(\psi^{\rm S}\big)=
\int_{L_M}\mathcal{D}\phi\, \psi^{\rm S}(q_{\Sigma}(\phi))O_M(\phi)e^{\mathrm{i} S (\phi)}.
%\label{eq:obsmapF}
\end{gather*}
We see from this expression that observables are considered as functions on spacetime
conf\/igu\-ra\-tion space in the Feynman quantization prescription.
Instead, in the Berezin--Toeplitz quantization prescription, observables are functions on phase
space.
The corresponding observable map, in the holomorphic representation, for the observable $O$ for
a~state $\psi^{\rm h}\in\mathcal{H}_{\Sigma}^{\rm h}$ is given as
\begin{gather}
\rho_M^{\blacktriangleleft O_M\blacktriangleright}\big(\psi^{\rm h}\big)=
\int_{L_{\tilde{M}}}\psi^{\rm h}(\xi)O_M(\xi)\mathrm{d}\nu_{\tilde{M}}(\xi),
\label{eq:obsmapBT}
\end{gather}
where $\xi\in L_\Sigma$ and $L_{\tilde{M}}$ is the space of solutions of the equation of motion,
def\/ined in a~neighborhood of the boundary hypersurface $\Sigma$ that admit a~well def\/ined
extension in the interior of the region~$M$.
$\mathrm{d}\nu_{\tilde{M}}(\xi)$ is a~suitable measure on $L_{\tilde{M}}$ and we refer
to~\cite{Oe:hol,Oe:Obs,Oe:SFobs} for details concerning the def\/inition of such structures.
In~\cite{Oe:SFobs} Oeckl was able to quantify the dif\/ference between the observable maps computed
in the Feynman quantization scheme and the one computed in the Berezin--Toeplitz quantization
scheme.
The result is presented in two propositions, in particular Propositions~4.3 and~4.7 of
the cited paper, where the amplitude of a~Weyl observable is derived for the two quantization
prescriptions.
We reproduce in the following formulas the statements of these propositions: For a~coherent state~$K_{\tau}$ we have
\begin{gather}
\rho^{\rm F}_M(K_{\tau})=
\rho_M(K_{\tau})F(\hat{\tau})\exp
\left(\frac{\mathrm{i}}{2}\int\mathrm{d}^2x\,\mu(x)\eta_D(x)-\frac{1}{2}g_{\Sigma}(\eta_D,\eta_D)\right),
\label{eq:obsmapWeyF}
\end{gather}
from the Feynman quantization (where $\hat{\tau}$ is a~complex solution of the equation of motion
determined by the coherent state $K_{\tau}$ and $\eta_D$ is the unique element of
$J_{\Sigma}L_{\tilde{M}}$ fulf\/illing the condition $D(\xi)=2\omega_{\partial M}(\xi,\eta_D)$ for
all $\xi\in L_{\tilde M}$, $\Sigma$ being the boundary of the region $M$) and
\begin{gather}
\rho^{\blacktriangleleft F\blacktriangleright}_M(K_{\tau})=
\rho_M(K_{\tau})F(\hat{\tau})\exp\left(-g_{\Sigma}(\eta_D,\eta_D)\right),
\label{eq:obsmapWeyBT}
\end{gather}
from the Berezin--Toeplitz quantization.

In particular the Feynman quantization prescription and the Berezin--Toeplitz quantization
prescription dif\/fer in a~property called composition correspondence~\cite{Oe:Obs}: while the
application of the observable axioms of the GBF to the product of classical observables with
disjoint support quantized via the Feynman quantization prescription leads to another observable
such that its expectation value is exactly the product of the expectation values of the original
observables (composition correspondence), this is not the case for the Berezin--Toeplitz
quantization prescription.

The following investigations will show another dif\/ference in the two quantization prescriptions:
First, we will show that the expectation value of the local observable $F(\phi)$ of the
form~\eqref{eq:Weylobs} in the state $D$ in Rindler space that corresponds to the thermal state
known from the derivation of the Unruh ef\/fect in the standard formulation of quantum f\/ield
theory coincides with the expectation value of $F(\phi)$ in the Minkowski vacuum when $F(\phi)$ is
localized in the interior of the right Rindler wedge.
Second, we will show that this is not true for the Berezin--Toeplitz quantization prescription.

\subsection{Expectation values in the Schr\"odinger representation}

\subsubsection{Observable maps from Feynman quantization}
\label{sec:obsmapSc}

Consider the spacetime region $M$ def\/ined in Section~\ref{sec:Min} in Minkowski spacetime.
We start by computing the observable amplitude
$\rho^{\rm F}_M:\mathcal{H}_{\Sigma_1}\otimes\mathcal{H}_{\overline{\Sigma}_2}\rightarrow{\mathbb C}$ for
the Weyl observable~\eqref{eq:Weylobs} on the quantum state, in the Schr\"odinger representation,
given by the tensor product of two copies of the vacuum state~\eqref{eq:vacMin}, namely
$K_{0,\Sigma_1}^{\rm S}\otimes\overline{K_{0,\Sigma_2}^{\rm S}}$.
Using the expression in equation~\eqref{eq:obsmapWeyF} we arrive at
\begin{gather}
\rho^{\rm F}_{M}\big(K_{0,\Sigma_1}^{\rm S}\otimes\overline{K_{0,\Sigma_2}^{\rm S}}\big)=
\exp\left(\frac{\mathrm{i}}{2}\int\mathrm{d}^2x\,\mathrm{d}^2x'\,\mu(x)G_{\rm F}^{\mathcal M}(x,x')\mu(x')\right),
\label{eq:expvalM1}
\end{gather}
where $G_{\rm F}^{\mathcal M}$ is the Feynman propagator in Minkowski spacetime, which is evaluated only
in the interior of the right Rindler wedge since the f\/ield $\mu(x)$ has support there.
The explicit form of the Feynman propagator can be obtained in terms of the expression of the boost
modes~\eqref{eq:boostmodes} in the right Rindler wedge, namely~\cite{BaEr:HigherTranFunc}
\begin{gather*}
\psi_k(x,t)\big|_{(x,t)\in\stackrel{\circ}{{\mathcal R}}}=
\frac{1}{\pi\sqrt{2}}\exp\left(\frac{\pi k}{2}-\mathrm{i}\frac{k}{2}
\ln\left(\frac{x+t}{x-t}\right)\right)K_{\mathrm{i}k}\big(m\sqrt{x^2-t^2}\big),
\end{gather*}
where $\stackrel{\circ}{{\mathcal R}}$ denotes the interior of the right Rindler wedge.
Then the Feynman propagator reads
\begin{gather}
G_{\rm F}^{\mathcal M}(x,x')\big|_{x,x' \in \stackrel{\circ}{{\mathcal R}}}  = \mathrm{i} \int_0^{\infty}
\frac{\mathrm{d} k}{\pi^2} \left\{ \cosh(\pi k) \cos\left( \frac{k}{2} \left( \ln \left(
\frac{x+t}{x-t}\right) - \ln \left( \frac{x'+t'}{x'-t'}\right) \right) \right) \right.
\nonumber
\\
\hphantom{G_{\rm F}^{\mathcal M}(x,x')\big|_{x,x' \in \stackrel{\circ}{{\mathcal R}}}  =}{}
 - \mathrm{i} \theta(t'-t) \sinh(\pi k) \sin\left( \frac{k}{2} \left( \ln \left(
\frac{x'+t'}{x'-t'}\right) - \ln \left( \frac{x+t}{x-t}\right) \right) \right)
\nonumber
\\
  \left.
\hphantom{G_{\rm F}^{\mathcal M}(x,x')\big|_{x,x' \in \stackrel{\circ}{{\mathcal R}}}  =}{}
- \mathrm{i} \theta(t'-t) \sinh(\pi k) \sin\left( \frac{k}{2} \left( \ln \left(
\frac{x+t}{x-t}\right) - \ln \left( \frac{x'+t'}{x'-t'}\right) \right) \right) \right\}
\nonumber
\\
\hphantom{G_{\rm F}^{\mathcal M}(x,x')\big|_{x,x' \in \stackrel{\circ}{{\mathcal R}}}  =}{}
 \times K_{\mathrm{i} k}(m \sqrt{x^2-t^2}) K_{\mathrm{i} k}(m \sqrt{{x'}^2-{t'}^2}).
\label{eq:FeyproMinR}
\end{gather}

Now, consider the region $R$ def\/ined in Section~\ref{sec:Rin} in Rindler spacetime.
The evaluation of the observable map is now performed on the state
$D\in\mathcal{H}_{\Sigma_1^R}\otimes\mathcal{H}_{\overline{\Sigma}_2^R}$ given by expression
\begin{gather}
D=\prod_i N_i^2\sum_{n_i=
0}^{\infty}\frac{e^{-2\pi n_i k_i/a}}{(n_i)!(2k_i)^{n_i}}\psi_{n_i}\otimes\overline{\psi_{n_i}},
\label{eq:thermalstate}
\end{gather}
where $\psi_{n_i}$ is the state with $n_i$ particles def\/ined in $\mathcal{H}_{\Sigma_i^R}$,
$i=1,2$,\footnote{Notice that the factor $(2k_i)^{n_i}$ appearing in the denominator    %   $(i=1,2)$ ->  $i=1,2$
of~\eqref{eq:thermalstate} comes from the normalization of the $n_i$-particle state,
\begin{gather*}
\int\mathcal{D}\varphi\,\psi_{k_1,\dots,k_n}(\varphi)\overline{\psi_{k_1',\dots,k_n'}(\varphi)}=
\frac{1}{n!}\sum_{\sigma\in S_n}\prod_{i=1}^n k_i\,\delta(k_i-k'_{\sigma(i)}),
\end{gather*}
where the sum runs over all permutations $\sigma$ of $n$ elements.}
and $N_i=\sqrt{1-\exp(-2\pi k_i/a)}$.
%and $N_i=(1-\exp(-2\pi k_i/a))^{\frac12}$.
In particular, the state $D$ corresponds to the mixed thermal state known from the derivation of
the Unruh ef\/fect in the standard formulation of quantum f\/ield theory~\cite{Cri:Unruh}.

From now on we set $a=1$.
Since for the observable map evaluated on coherent states we can use the general result in
equation~\eqref{eq:obsmapWeyF}\footnote{See also~\cite{CoDo:gen} for the expression of amplitude
maps in terms of modes expansion.}, it is convenient to express the state in
equation~\eqref{eq:thermalstate} in terms of coherent states\footnote{An important property
satisf\/ied by coherent states is the completeness relation expressed by the resolution of the
identity operator id which, in a~bra ket notation, takes the form
\begin{gather*}
N^{-1}\int\mathrm{d}\xi\,\mathrm{d}\overline{\xi}\,|K^{\rm S}_{\xi}\rangle\langle K^{\rm S}_\xi|=\text{id},
%\label{eq:csresid}
\qquad
\text{with}
\qquad
N=\int\mathrm{d}\xi\,\mathrm{d}\overline{\xi}\,\exp\left(-\int\frac{\mathrm{d}k}{2k}\,|\xi(k)|^2\right).
\end{gather*}}; the observable map in the region $R$ for the state $D$ then reads
\begin{gather}
\rho_R^{\rm F}(D)=\prod_i N_i^2\sum_{n_i=0}^{\infty}
\frac{e^{-2\pi n_i k_i}}{(n_i)!(2k_i)^{n_i}}\,N^{-2}
\int\mathrm{d}\xi_1\,\mathrm{d}\overline{\xi_1}\,\mathrm{d}\xi_2\,\mathrm{d}\overline{\xi_2}\,\rho^{\rm F}_R
\big(K^{\rm S}_{\xi_1}\otimes\overline{K^{\rm S}_{\xi_2}}\big)
\nonumber
\\
\phantom{\rho_R^{\rm F}(D)=}{}
\times
\exp\left(-\frac{1}{2}\int\frac{\mathrm{d}k}{2k}|\xi_1(k)|^2\right)
(\xi_1(k_i))^{n_i}\,\exp\left(-\frac{1}{2}\int\frac{\mathrm{d}k}{2k}|\xi_2(k)|^2\right)(\overline{\xi_2(k_i)})^{n_i},
\label{eq:obsmapR}
\end{gather}
where the terms in the second line come from the scalar product of the $n_i$-particle states
appearing in~\eqref{eq:thermalstate} and the coherent states $K^{\rm S}_{\xi_1}$
and $\overline{K^{\rm S}_{\xi_2}}$ respectively, see Section~II.B of~\cite{CoOe:Smatrix}.
The observable map $\rho^{\rm F}_R(K^{\rm S}_{\xi_1}\otimes\overline{K^{\rm S}_{\xi_2}})$ has been shown to satisfy
a~factorization property, see Proposition~4.3 of~\cite{Oe:SFobs}, which corresponds to the
amplitude map of the theory with a~source f\/ield interaction~\cite{CoDo:gen},
\begin{gather}
\rho^{\rm F}_R\big(K^{\rm S}_{\xi_1}\otimes\overline{K^{\rm S}_{\xi_2}}\big)=\rho_R\big(K^{\rm S}_{\xi_1}\otimes\overline{K^{\rm S}_{\xi_2}}\big)
\exp\left(\int\mathrm{d}^2x\,\hat{\xi}(x)\mu(x)\right)
\nonumber
\\
\phantom{\rho^{\rm F}_R\big(K^{\rm S}_{\xi_1}\otimes\overline{K^{\rm S}_{\xi_2}}\big)=}
\times\exp
\left(\frac{\mathrm{i}}{2}\int\mathrm{d}^2x\,\mathrm{d}^2x'\,\mu(x)G_{\rm F}^{\mathcal R}(x,x')\mu(x')\right),
\label{eq:expval02}
\end{gather}
where we are now using $x$ as global notation for the Rindler coordinates $(\eta,\rho)$.
The f\/irst term in the r.h.s.\
of~\eqref{eq:expval02} is the free amplitude map~\eqref{eq:freampmap} for the state
$K^{\rm S}_{\xi_1}\otimes\overline{K^{\rm S}_{\xi_2}}$,
\begin{gather*}
\rho_R\big(K^{\rm S}_{\xi_1}\otimes\overline{K^{\rm S}_{\xi_2}}\big)=
\exp\left(\int_0^{\infty}\frac{\mathrm{d}k}{2k}
\left(\xi_1(2)\overline{\xi_2(k)}-\frac{1}{2}|\xi_1(k)|^2-\frac{1}{2}|\xi_2(k)|^2\right)\right).
\end{gather*}
$\hat{\xi}(x)$ is a~complex solution of the equation of motion determined by the two coherent
states $K^{\rm S}_{\xi_1}$ and $\overline{K^{\rm S}_{\xi_2}}$,
\begin{gather*}
\hat{\xi}(x)=
\mathrm{i}\int_0^{\infty}\mathrm{d}k\left(\phi^R_k(x)\,\xi_1(k)+\overline{\phi^R_k(x)}\,\overline{\xi_2(k)}\right),
%\label{eq:xihat}
\end{gather*}
where $\phi^R_k(\rho,\eta)$ are the Fulling modes~\eqref{eq:Rinmodes}.
Finally, $G_{\rm F}^{\mathcal R}(x,x')$ appearing in the last term of~\eqref{eq:expval02} is the Feynman
propagator in Rindler spacetime and in the region $R$ it reads\footnote{The general expression of
the Feynman propagator for f\/ields in (a wide class of) curved spacetimes has been obtained
in~\cite{CoDo:gen}, to which we refer also for details concerning the calculation presented here.}
\begin{gather}
G_{\rm F}^{\mathcal R}(\rho,\eta,\rho',\eta')=
\mathrm{i}\int_0^{\infty}\frac{\mathrm{d}k}{\pi^2}
\left(\theta(\eta'-\eta)e^{-\mathrm{i}k(\eta'-\eta)}+\theta(\eta-\eta')e^{-\mathrm{i}k(\eta-\eta')}\right)
\nonumber
\\
\phantom{G_{\rm F}^{\mathcal R}(\rho,\eta,\rho',\eta')=}
{} \times
K_{\mathrm{i}k}(m\rho)K_{\mathrm{i}k}(m\rho')\sinh(\pi k).
\label{eq:FeyproRin}
\end{gather}
We now have at our disposal all the ingredients to compute the integrals in~\eqref{eq:obsmapR}.
It is convenient to proceed by expressing the powers of the modes $\xi_{1,2}(k_i)$
in~\eqref{eq:obsmapR} in terms of functional derivatives,
\begin{gather*}
(\xi_1(k_i))^{n_i} (\overline{\xi_2(k_i)})^{n_i}=
(2k_i)^{2n_i}\frac{\delta^{n_i}}{\delta\alpha(k_i)^{n_i}}\frac{\delta^{n_i}}{\delta\beta(k_i)^{n_i}}\,
e^{\left(\int\frac{\mathrm{d}k}{2k}\left(\beta(k)\xi_1(k)+\alpha(k)\overline{\xi_2(k)}\right)\right)}
%\exp\left(\int\frac{\mathrm{d}k}{2k}\left(\beta(k)\xi_1(k)+\alpha(k)\overline{\xi_2(k)}\right)\right)
\bigg|_{\alpha=\beta=0}.
\end{gather*}
Substituting this in~\eqref{eq:obsmapR}, the integrals are evaluated by the following shift of
integration va\-riables:
\begin{alignat*}{3}
&\xi_1 \rightarrow \xi_1 + \beta + \mu_1,\qquad & &\overline{\xi_1} \rightarrow \overline{\xi_1} + \alpha,&
\\
&\xi_2 \rightarrow \xi_2 + \alpha,\qquad & &\overline{\xi_2} \rightarrow \overline{\xi_2} + \overline{\xi_1} + \mu_2,&
\end{alignat*}
where $\mu_1(k)=\mathrm{i}\int\mathrm{d}x^2\,2k\,\phi^R_k(x)\mu(x)$
and $\mu_2(k)=\mathrm{i}\int\mathrm{d}x^2\,2k\overline{\phi^R_k(x)}\mu(x)$.
We arrive at
\begin{gather*}
\rho_R^{\rm F}(D)  = \prod_i N_i^2 \sum_{n_i=0}^{\infty} \frac{e^{- 2 \pi n_i k_i}}{(n_i)!}
(2k_i)^{n_i} \frac{\delta^{n_i}}{\delta \alpha(k_i)^{n_i}} \frac{\delta^{n_i}}{\delta
\beta(k_i)^{n_i}}
\\
\hphantom{\rho_R^{\rm F}(D)  =}{}
\times\exp \left( \int \frac{\mathrm{d} k}{2k} \left( \beta(k) \mu_1(k) + \alpha(k) \mu_2(k) +
\alpha(k) \beta(k)\right) \right) \bigg|_{\alpha=\beta=0}
\\
\hphantom{\rho_R^{\rm F}(D)  =}{}
 \times \exp \left( \frac{\mathrm{i}}{2} \int \mathrm{d}^2 x \, \mathrm{d}^2 x' \, \mu(x)
G_{\rm F}^{\mathcal R}(x,x') \mu(x')\right),
\\
\hphantom{\rho_R^{\rm F}(D)}{}
= \prod_i N_i^2 \sum_{n_i=0}^{\infty} \frac{e^{- 2 \pi n_i k_i}}{(n_i)!}
\frac{\delta^{n_i}}{\delta \alpha(k_i)^{n_i}} (\alpha(k_i) + \mu_1(k_i))^{n_i}   \exp \left( \int
\frac{\mathrm{d} k}{2k} \, \alpha(k) \mu_2(k) \right) \bigg|_{\alpha=0}
\\
\hphantom{\rho_R^{\rm F}(D)  =}{}
 \times \exp \left( \frac{\mathrm{i}}{2} \int \mathrm{d}^2 x \, \mathrm{d}^2 x' \, \mu(x)
G_{\rm F}^{\mathcal R}(x,x') \mu(x')\right).
%\label{eq:obsmapR2}
\end{gather*}
To compute the derivative with respect to $\alpha$ we use Rodrighues' formula, see~\cite[(8.970.1)]{GrRi}, and obtain
\begin{gather*}
\rho_R^{\rm F} (D)=\prod_i \!N_i^2\!\sum_{n_i=0}^{\infty}\!e^{-2\pi n_i k_i}L_{n_i}
\left(\!-\frac{\mu_1(k_i)\mu_2(k_i)}{2k_i}\right)
\exp\left(\frac{\mathrm{i}}{2}\int\mathrm{d}^2x\,\mathrm{d}^2x'\,\mu(x)G_{\rm F}^{\mathcal R}(x,x')\mu(x')\right)\!,
\end{gather*}
where $L_{n_i}$ is the Laguerre polynomial of order $n_i$.
According to formula~(8.975.1) of~\cite{GrRi}, the sum over $n_i$ gives
\begin{gather*}
\rho^{\rm F}_R(D)=
\prod_i \exp\left(\frac{\mu_1(k_i)\mu_2(k_i)}{2k_i}\frac{e^{-\pi k_i}}{2\sinh(\pi k_i)}\right)
\exp\left(\frac{\mathrm{i}}{2}\int\mathrm{d}^2x\,\mathrm{d}^2x'\,\mu(x)G_{\rm F}^{\mathcal R}(x,x')\mu(x')\right).
\end{gather*}
Finally, the substitution of the expression of the quantities $\mu_1(k_i)$ and $\mu_2(k_i)$ leads to
\begin{gather}
\rho^{\rm F}_R(D)=
\exp\left(\frac{\mathrm{i}}{2}\int\!\mathrm{d}^2x\,\mathrm{d}^2x'\,\mu(x)
\left[\mathrm{i}\int\!\mathrm{d}k\,\phi^R_k(x)\overline{\phi^R_k(x')}
\frac{e^{-\pi k}}{\sinh(\pi k)}+G_{\rm F}^{\mathcal R}(x,x')\right]\mu(x')\right).\!\!
\label{eq:obsmapR3}
\end{gather}
Noticing that only the symmetric component of the f\/irst term in the square bracket contributes to
the integral and using~\eqref{eq:Rinmodes} and~\eqref{eq:FeyproRin}, a~straightforward calculation
shows that the sum in the square bracket coincides with the Feynman propagator in Minkowski
spacetime evaluated in the right Rindler wedge~\eqref{eq:FeyproMinR}; and so do the observable
maps~\eqref{eq:expvalM1} and~\eqref{eq:obsmapR3}.
In the next section we present the same calculation performed according to the Berezin--Toeplitz
quantization~\eqref{eq:obsmapBT} of the Weyl observable.

\subsubsection[Observable maps from Berezin-Toeplitz quantization]{Observable maps from Berezin--Toeplitz quantization}
\label{sec:expvalBT}

By examining the expressions~\eqref{eq:obsmapWeyF} and \eqref{eq:obsmapWeyBT} one can see that the
dif\/ference between the obser\-vable maps of a~Weyl observable in the two quantization schemes
amounts to an exponential factor.
Moreover the exponential factor in~\eqref{eq:obsmapWeyF} corresponds to the last exponential factor
in~\eqref{eq:expval02},
and consequently we have that
\begin{gather*}
-g_{\Sigma}(\eta_D,\eta_D)=-\int\mathrm{d}^2x\,\mathrm{d}^2x'\,\mu(x) \Im(G_{\rm F}(x,x'))\,\mu(x').
\end{gather*}
Hence, in Minkowski spacetime the observable map~\eqref{eq:obsmapWeyBT} on the vacuum state
evaluated in the Berezin--Toeplitz quantization scheme in the spacetime region $M$, is given by
\begin{gather}
\rho^{\blacktriangleleft F\blacktriangleright}_M(\psi_0\otimes\overline{\psi_0})=
\exp\left(-\int\mathrm{d}^2x\,\mathrm{d}^2x'\,\mu(x)\,\Im\big(G_{\rm F}^M(x,x')\big) \mu(x')\right).
\label{eq:expvalBTM}
\end{gather}
In Rindler spacetime the observable map~\eqref{eq:obsmapWeyBT}, in the same quantization scheme in
the spacetime region $R$, takes the form
\begin{gather}
\rho^{\blacktriangleleft F\blacktriangleright}_R(D)=
\exp\Bigg({-}\int\mathrm{d}^2x\,\mathrm{d}^2x'\,\mu(x)
\nonumber
\\
\phantom{\rho^{\blacktriangleleft F\blacktriangleright}_R(D)=}
{}\times
\left[\frac{1}{2}\int\mathrm{d}k\,\phi^R_k(x)\overline{\phi^R_k(x')}\,
\frac{e^{-\pi k}}{\sinh(\pi k)}+\Im\left(G^R_{\rm F}(x,x')\right)\right]\mu(x')\Bigg).
\label{eq:expvalBTR}
\end{gather}
As in~\eqref{eq:obsmapR3} only the symmetric part of the terms in the square bracket contribute to
the integral and the situation is similar to the one in the preceding section apart from the factor~$1/2$ appearing in~\eqref{eq:expvalBTR}.
It is precisely this factor that prevents the coincidence of~\eqref{eq:expvalBTM}
and~\eqref{eq:expvalBTR}.
We conclude that the Berezin--Toeplitz prescription for the quantization of observables gives no
ground for the Unruh ef\/fect.

\subsection{Expectation values in the holomorphic representation}

In this section, we present the computation of the observable maps~\eqref{eq:obsmapWeyF}
and \eqref{eq:obsmapWeyBT} for quantum states in the holomorphic
representation\footnote{In~\cite{Oe:Sch-hol} a~one-to-one relation was established between the
Schr\"odinger and the holomorphic representation in terms of an isomorphism between the
corresponding Hilbert spaces.
Thus, by using this result it will be possible to obtain the amplitude and observable maps in the
holomorphic representation starting from those obtained in the Schr\"odinger representation.
We shall elaborate on this elsewhere and follow here a~dif\/ferent strategy: We start with the
state~\eqref{eq:mixstahol} and compute the observable map of the Weyl observable with the
prescription suited for the holomorphic representation.}.
First we notice that in Minkowski spacetime~\eqref{eq:obsmapWeyF} for the vacuum state reduces to
the same result obtained in Section~\ref{sec:obsmapSc}, namely expression~\eqref{eq:expvalM1}; this is a~consequence of the equivalence
between the Schr\"odinger and holomorphic quantizations shown in~\cite{Oe:Sch-hol}.
We now consider the same observable map in Rindler spacetime on the state $D^{\rm h}$ in the holomorphic
representation corresponding to the state in equation~\eqref{eq:thermalstate} in the Schr\"odinger
representation.
For later convenience we write this state in terms of derivatives of coherent states,
\begin{gather}
D^{\rm h}=\left.
\prod_k N^2_k\sum_{n=0}^\infty e^{-2\pi n k}\frac{2^{n}}{n!}
\frac{\delta^n}{\delta\xi_{1}(k)^n}
\frac{\delta^n}{\delta\xi_{2}(k)^n} K^{\rm h}_{\xi_1}\otimes\overline{K^{\rm h}_{\xi_2}}\right|_{\xi_1=\xi_2=0},
\label{eq:mixstahol}
\end{gather}
where $K^{\rm h}_{\xi_1}\in\mathcal{H}^{\rm h}_{\Sigma_1}$ and $K^{\rm h}_{\xi_2}\in\mathcal{H}^{\rm h}_{\Sigma_2}$ are
the coherent states in the holomorphic representation def\/ined by $\xi_{i}\in L_{\Sigma_i^R}$,
$i=1,2$.
Consequently $D^{\rm h}$ is a~state in the whole boundary Hilbert space
$\mathcal{H}^{\rm h}_{\Sigma_1}\otimes\mathcal{H}^{\rm h}_{\overline{\Sigma}_2}$.
The corresponding observable map for the Weyl observable~\eqref{eq:Weylobs} reads
\begin{gather}
\rho^{\rm F}_R\big(D^{\rm h}\big)=\left.
\prod_k N^2_k\sum_{n=0}^\infty e^{-2\pi n k}\frac{2^{n}}{n!}
\frac{\delta^n}{\delta\xi_{1}(k)^n}
\frac{\delta^n}{\delta\xi_{2}(k)^n}\rho^{\rm F}_R\big(K^{\rm h}_{\xi_1}\otimes\overline{K^{\rm h}_{\xi_2}}\big)
\right|_{\xi_1=\xi_2=0}.
\label{eq:expvalhol1}
\end{gather}
We now specify the three terms appearing in the expression~\eqref{eq:obsmapWeyF} for the observable
map $\rho^{\rm F}_R(K^{\rm h}_{\xi_1}\otimes\overline{K^{\rm h}_{\xi_2}})$:
\begin{itemize}\itemsep=0pt
\item the free amplitude $\rho_R(K^{\rm h}_{\xi_1}\otimes\overline{K^{\rm h}_{\xi_2}})$ can be computed
using~\eqref{eq:freeamphol}, where in the present context $\xi^R=\xi_1+\xi_2$
and $\xi^I=\xi_1-\xi_2$, leading to
\begin{gather*}
\rho_R\big(K^{\rm h}_{\xi_1}\otimes\overline{K^{\rm h}_{\xi_2}}\big)=
\exp\left(\frac{1}{2}\int_0^{\infty}\mathrm{d}k\,\xi_{1}(k)\xi_{2}(k)\right),
\end{gather*}
\item the Weyl observable evaluated on the complex solution $\hat{\xi}$ given in this case
by\footnote{As in the previous section $x$ is used as global notation for the Rindler coordinates
$(\rho,\eta)$.}
\begin{gather*}
\hat{\xi}(x)=\xi^R(x)-\mathrm{i}\xi^I(x)=
\frac{1}{\sqrt{2}}\int_0^{\infty}\mathrm{d}k\left(\phi_k^R(x)\xi_1(k)+\overline{\phi_k^R(x)}\xi_2(k)\right),
\end{gather*}
\item the last term in the r.h.s.\ of~\eqref{eq:obsmapWeyF} coincides with the last term in the r.h.s.\
of~\eqref{eq:expval02}.
\end{itemize}
The observable map~\eqref{eq:expvalhol1} can then be written as
\begin{gather}
\nonumber
\rho^{\rm F}_R\big(D^{\rm h}\big)= \prod_k N^2_k\sum_{n=0}^\infty e^{-2\pi n k} \frac{2^{n}}{n!}
\frac{\delta^n}{\delta \xi_{1}(k)^n} \frac{\delta^n}{\delta \xi_{2}(k)^n}
\nonumber
\\
\left.
\hphantom{\rho^{\rm F}_R(D^{\rm h})=}{}
 \times
\exp \left(\frac{1}{2} \, \xi_{1}(k) \xi_{2}(k) + \frac{\mathrm{i}}{\sqrt{2}} \int \mathrm{d}^2x \,
\mu(x) \left( \phi_k^R(x) \xi_1(k) + \overline{\phi_k^R(x)} \xi_2(k) \right)\right)
\right|_{\xi_1=\xi_2=0}
\nonumber
\\
\hphantom{\rho^{\rm F}_R(D^{\rm h})=}{}
 \times \exp \left( \frac{\mathrm{i}}{2} \int \mathrm{d}^2 x \, \mathrm{d}^2 x' \, \mu(x)
G_{\rm F}^{\mathcal R}(x,x') \mu(x') \right).
\label{eq:expvalhol2}
\end{gather}
We proceed by evaluating the f\/irst line in the r.h.s.\
of~\eqref{eq:expvalhol2} by applying the general Leibniz rule
\begin{gather*}
\frac{\mathrm{d}^n}{\mathrm{d}\gamma^n}f(\gamma)g(\gamma)=\sum_{k=0}^{n}\binom{n}{k}
\frac{\mathrm{d}^{n-k}}{\mathrm{d}\gamma^{n-k}}f(\gamma)\frac{\mathrm{d}^{k}}{\mathrm{d}\gamma^{k}}g(\gamma),
\end{gather*}
and using the relation
\begin{gather*}
\sum_{k=0}^{\infty}\frac{(k+s)!}{k!s!}e^{-2\pi kp}=\frac{1}{(1-e^{-2\pi p})^{s+1}},
\end{gather*}
which we proof in the appendix.
We obtain
\begin{gather}
\nonumber
N_k^2\sum_{n=0}^\infty
 e^{-2\pi n k} \frac{2^{n}}{n!} \frac{\delta^n}{\delta \xi_{1}(k)^n}
\frac{\delta^n}{\delta \xi_{2}(k)^n}\exp\left(\frac{1}{2} \, \xi_{1}(k) \xi_{2}(k) \right) \times
\\
\nonumber
\qquad\quad{} \times\left.\exp\left( \frac{\mathrm{i}}{\sqrt{2}} \int \mathrm{d}^2x \, \mu(x) \left( \phi_k^R(x)
\xi_1(k) + \overline{\phi_k^R(x)} \xi_2(k) \right)\right) \right|_{\xi_1=\xi_2=0}
\\
\nonumber
\qquad{} = N_k^2 \sum_{n=0}^\infty e^{-2\pi n k} \left(-  \int \mathrm{d}^2x \, \mathrm{d}^2x' \, \mu(x)
\mu(x') \phi_k^R(x) \overline{\phi_k^R(x')}\right)^{n} \frac{1}{n!} \sum_{j=0}^{\infty}
\frac{(j+n)!}{j!n!} e^{-2\pi kj}
\\
\qquad{} = \exp \left(-\frac{e^{-\pi k}}{2 \sinh(\pi k)} \int \mathrm{d}^2x \, \mathrm{d}^2x' \, \mu(x)
\mu(x') \phi_k^R(x) \overline{\phi_k^R(x')} \right).
\end{gather}
Hence, substituting in~\eqref{eq:expvalhol2} we obtain after some rearrangements
\begin{gather}
\rho_R^{\rm F}\big(D^{\rm h}\big)=
\exp\left(\frac{\mathrm{i}}{2}\!\int\!\mathrm{d}^2x\,\mathrm{d}^2x'\,\mu(x)\!
\left[\mathrm{i}\!\int\!\mathrm{d}k\,\phi_k^R(x)\overline{\phi_k^R(x')}\frac{e^{-\pi k}}{\sinh(\pi k)}
+G_{\rm F}^{\mathcal R}(x,x')\right]\!\mu(x')\right)\!,\!\!\!
\label{eq:obsmaphol3}
\end{gather}
which coincides with expression~\eqref{eq:obsmapR3}.
Consequently,~\eqref{eq:obsmaphol3} equals the observable map computed in Minkowski spacetime on
the vacuum state, and we recover the coincidence of the expectation values we found for the
Schr\"odinger representation also in the holomorphic representation.

As already noticed, the dif\/ference between the Berezin--Toeplitz quantization and the Feynman one
amounts to the last factors in~\eqref{eq:obsmapWeyF} and \eqref{eq:obsmapWeyBT}.
These terms are independent of the representation chosen for the quantum states, and so we are
reduced to the same situation as in Section~\eqref{sec:expvalBT}: the coincidence of expectation values that exists for the Feynman
quantization prescription does not appear adopting the Berezin--Toeplitz prescription for
quantizing local observables.

\section{Conclusions and outlook}
\label{sec:con}

We have applied the general boundary formulation of quantum f\/ield theory to quantize a~massive
scalar f\/ield in Minkowski and Rindler spacetimes.
We showed that the expectation values of Weyl observables with compact spacetime support in the
interior of the right Rindler wedge, computed in the Minkowski vacuum state coincide with those
calculated in a~state in Rindler space that corresponds to the thermal state known from the
derivation of the Unruh ef\/fect in the standard formulation of quantum f\/ield theory if the
observables are quantized according to the Feynman quantization prescription.
This result could be interpreted as the manifestation of the Unruh ef\/fect within the GBF.
Furthermore, we showed that the coincidence of the expectation values does not hold in the
Berezin--Toeplitz quantization, which is an alternative quantization scheme for observables in the
GBF.

The work in this article is of immediate relevance for the GBF program.
It represents a~concrete application of the quantization of observables and the opportunity to
compare the Feynman quantization prescription and Berezin--Toeplitz quantization prescription in
a~specif\/ic context.
The observed dif\/ference in the two quantization prescriptions adds up to the dif\/ference in the
existence of a~property called composition correspondence~\cite{Oe:Obs}: while the application of
the observable axioms of the GBF to the product of classical observables with disjoint support
quantized via the Feynman quantization prescription leads to another observable such that its
expectation value is exactly the product of the expectation values of the original observables,
this is not the case for the Berezin--Toeplitz quantization prescription.
It will be of interest for the GBF to explore the dif\/ferences of the two quantization
prescriptions in more detail.

Another issue for future work is the def\/inition of KMS states in the GBF.
Such a~def\/inition could be used to make the relation between the quantum theory in Minkowski
and the one in the Rindler wedge more explicit.
It would be interesting to explore if the corresponding results of the GBF would dif\/fer from
those found in algebraic quantum f\/ield theory.

It should be noted that the spacetime regions considered for the evaluation of the observable maps
are the standard ones bounded by two equal (Minkowski and Rindler) time hyperplanes.
Of course the versatility of GBF enables to quantize the f\/ield and to compute expectation values
in more general regions.
Although the main focus is represented by compact spacetime regions, inspired by previous results
obtained applying the GBF in Minkowski and curved spacetime, an interesting region is represented
by the one bounded by one connected and timelike boundary.
In particular, it is possible to apply the GBF for a~f\/ield def\/ined in a~region of Rindler
spacetime bounded by one hyperbola of constant Rindler spatial coordinate $\rho$.
The origin of Minkowski spacetime lies outside this region and the comparison of the quantum
f\/ield theory def\/ined there and the one in Minkowski will then avoid the dif\/f\/iculty inherent
with the behavior of the f\/ield in $(x,t)=(x,0)$.

Furthermore, the analysis of the properties of the Minkowski and Rindler quantum theories can be
the basis for solving an open question within the GBF.
The hyperplane $t=0$ in Minkowski spacetime is the union of the two semi-hyperplanes $\eta^R=0$
and $\eta^L=0$ in the right and left Rindler wedge respectively.
However the Hilbert space associated to the hypersurface $t=0$ is not the tensor product of the
Hilbert spaces associated with $\eta^R=0$ and $\eta^L=0$, due to the additional boundary condition
at the origin.
In order to compare the dif\/ferent Hilbert spaces one possibility would be to consider
hypersurfaces with boundaries: in the present context the hyperplane $t=0$ for $x\geq0$ ($x\leq0$),
namely with a~boundary in the origin of Minkowski spacetime.
However it is still not clear within the GBF which algebraic structure should be associated with an
hypersurface with boundaries\footnote{Oeckl~R., Private communication.}. %\cite{Oe:private}.
The solution of such a~question is of paramount importance from the perspective of the Unruh
ef\/fect in the GBF, as well as for more general contexts.

\appendix

\section{Appendix}

Here we prove the identity
\begin{gather*}
\sum_{k=0}^{\infty}\frac{(k+n)!}{k!n!}e^{-2\pi kp/a}=\frac{1}{(1-e^{-2\pi p/a})^{n+1}}.
\end{gather*}
\begin{proof}
We start by remarking that with
\begin{gather*}
f(n):=\sum_{k=0}^\infty\frac{(k+n)!}{k!n!}e^{-2\pi kp/a}
\end{gather*}
we have
\begin{gather*}
f(n+1)=\left(1-\frac{1}{n+1}\frac{a}{2\pi}\frac{d}{dp}\right)f(n).
\end{gather*}
For $s=0$ we f\/ind
\begin{gather*}
f(0)=\frac{1}{1-e^{-2\pi p/a}}.
\end{gather*}
So we start the induction step with the ansatz
\begin{gather*}
f(n)=\frac{1}{(1-e^{-2\pi p/a})^{n+1}}
\end{gather*}
and f\/ind
\begin{gather*}
f(n+1)=
\left(1-\frac{1}{n+1}\frac{a}{2\pi}\frac{d}{dp}\right)\frac{1}{\big(1-e^{-2\pi p/a}\big)^{n+1}}
\\
\phantom{f(n+1)}
=\frac{1}{(1-e^{-2\pi p/a})^{n+1}}+\frac{e^{-2\pi p/a}}{(1-e^{-2\pi p/a})^{n+2}}=\frac{1}{(1-e^{-2\pi p/a})^{n+2}}
\end{gather*}
which proves that the ansatz was correct.
\end{proof}

\subsection*{Acknowledgements}
The authors thank Robert Oeckl for very helpful discussions and remarks.
The authors also thank the anonymous referees which gave a
relevant contribution to improve the paper.
Part of this work was done during a~research stay of DR at the UNAM Campus Morelia funded by
CONACYT grant 49093.
The work of DR has been supported by the International Max Planck Research School for Geometric
Analysis, Gravitation and String Theory.
The work of DC has been supported in part by UNAM--DGAPA--PAPIIT through project grant IN100212.

\pdfbookmark[1]{References}{ref}
\LastPageEnding

\end{document}